# Response of a tungsten powder target to an incident high energy proton beam

O. Caretta, T. Davenne, C. Densham, M. Fitton, P. Loveridge, and J. O'Dell

*STFC Rutherford Appleton Laboratory, Didcot OX11 0QX, United Kingdom*

N. Charitonidis, I. Efthymiopoulos, and A. Fabich

*CERN, EN-MEF, 1211 Geneva 23, Switzerland*

L. Rivkin

*Paul Scherrer Institut (PSI), CH-5232 Villigen PSI, Switzerland*

The experiment described in this paper is the first study of the response of a static tungsten powder sample to an impinging high energy proton beam pulse. The experiment was carried out at the HiRadMat facility at CERN. Observations include high speed videos of a proton beam induced perturbation of the powder sample as well as data from a laser Doppler vibrometer measuring the oscillations of the powder container. A comparison with a previous analogous experiment which studied a proton beam interaction with mercury is made.

## I. INTRODUCTION

A continuously flowing fluidized tungsten powder jet [1] has been proposed as a potential target technology for future high energy physics facilities (e.g., for a Neutrino Factory [2] or a compact neutron source). The Neutrino Factory concept requires a high power proton beam to interact with a target to generate mesons (pions) and eventually a beam of neutrinos through lepton (muon) decays. During the interaction significant heat is deposited in the target material which results in high temperatures and high stress. The interest in flowing tungsten powder technology arose from its potential to accommodate very high deposited power densities while maintaining a reasonable operating temperature and low stress levels.

The fluidization of tungsten powder has been demonstrated off-line in a bespoke test facility [3], using both air and helium as the carrier gas. Both open jets and contained flows of dense phase powder have been generated in a horizontal configuration suitable for a particle accelerator target system. Recirculation of the material in a combination of dense and lean phases has been achieved as required for a continuously operating facility. Further research to optimize this technology is currently under way.

This paper reports on the response of a stationary tungsten powder sample in helium to incident pulses from the 440 GeV proton beam at the HiRadMat facility at

CERN [4]. Similar experiments were conducted in the past on mercury since an open liquid mercury jet is the current baseline concept for a Neutrino Factory target [2]. Recent work shows that either tungsten powder or mercury could achieve a comparable neutrino yield when used as targets [5]. The first experiment with mercury involved a static mercury thimble [6] and a later experiment looked at the response of flowing mercury inside a magnetic field [7]. The stationary mercury thimble experiment measured beam induced splash velocities of the order of 30 m/s for a pulsed energy density of 28 J/g. The interest in measuring the perturbation velocity of the target material arises from the potential of high speed impacts being damaging to the surrounding containment. This experiment is in many ways analogous to the stationary mercury thimble experiment.

## II. EXPERIMENT DESCRIPTION/APPARATUS

The CERN HiRadMat facility has a test area specifically designed to facilitate the study of high intensity pulsed beam interactions with materials. At HiRadMat a high-energy 440 GeV/$c$ proton beam is extracted from the CERN super proton synchrotron (SPS) and transported a few hundred meters to the irradiation area where it can be focused onto the experimental targets. The number of bunches, bunch spacing, bunch intensity and beam spot size are adjustable in order to suit the needs of the experiments.

The position of the beam was monitored using a set of two button beam position monitors (BPM), located approximately 2 and 10 meters upstream of the experiment. The intensity of the extracted beam was measured using a fast beam current transformer, while the transverse beam profile was measured using fluorescence screens. The information from all monitors as well as from all magnets and settings of the SPS were logged in a database.

TABLE I. List of extracted beam pulses to the experiment. For the three empty shots, the beam was dumped into the SPS and not extracted to the experiment.

| Shot no. | Time | Bunches/pulse | Intensity protons on target [PoT] | Beam sigma $\sigma_x$ [mm] | Beam sigma $\sigma_y$ [mm] | LDV aim inner or outer trough | Camera frame rate [kHz] | Beam depth from trough lip [mm] |
|---|---|---|---|---|---|---|---|---|
| 1 | 12:55 | 1 | $6.80 \times 10^9$ | ⋯ | ⋯ | IN | 2 | 6 |
| 2 | 13:17 | 1 | $3.50 \times 10^9$ | ⋯ | ⋯ | OUT | 2 | 6 |
| 3 | ⋯ | No beam | ⋯ | | | | | |
| 4 | 13:43 | 6 | $4.60 \times 10^{10}$ | 0.45 | 1 | IN | 2 | 6 |
| 5 | 14:07 | 6 | $4.36 \times 10^{10}$ | 0.76 | 1.09 | OUT | 2 | 6 |
| 6 | 14:20 | 6 | $8.10 \times 10^{10}$ | 0.77 | 1.23 | IN | 2 | 6 |
| 7 | 14:25 | 6 | $8.04 \times 10^{10}$ | 0.8 | 1.2 | OUT | 2 | 6 |
| 8 | 14:41 | 6 | $1.75 \times 10^{11}$ | 1.17 | 1.6 | IN | 2 | 6 |
| 9 | 14:50 | 6 | $1.85 \times 10^{11}$ | 1.07 | 1.3 | OUT | 2 | 6 |
| 10 | 15:01 | 6 | $1.58 \times 10^{11}$ | 0.93 | 1.66 | IN | 1 | 6 |
| 11 | 15:08 | 6 | $1.69 \times 10^{11}$ | 1.1 | 1.6 | IN | 2 | 6 |
| 12 | 15:53 | 6 | $1.30 \times 10^{11}$ | 2.15 | 1.61 | IN | 2 | 6 |
| 13 | 15:58 | 6 | $1.60 \times 10^{11}$ | 2.14 | 1.69 | IN | 1 | 6 |
| 14 | 16:03 | 6 | $2.00 \times 10^{11}$ | 2.33 | 1.88 | OUT | 1 | 6 |
| 15 | ⋯ | No beam | ⋯ | | | | | |
| 16 | 17:04 | 36 | $1.49 \times 10^{11}$ | 1 | 1.51 | IN | 1 | 6 |
| 17 | 17:17 | 36 | $2.00 \times 10^{11}$ | 1.1 | 1.7 | IN | 1 | 6 |
| 18 | 17:21 | 36 | $2.60 \times 10^{11}$ | 1.37 | 1.79 | OUT | | 6 |
| 19 | ⋯ | No beam | ⋯ | | | | | |
| 20 | 17:33 | 36 | $2.64 \times 10^{11}$ | 1.31 | 1.81 | IN | 1 | 6 |
| 21 | 17:37 | 36 | $2.94 \times 10^{11}$ | 1.39 | 1.85 | IN | 1 | 6 |
| 22 | 17:57 | 36 | $1.58 \times 10^{11}$ | 0.92 | 1.66 | IN | 1 | 4 |
| 23 | ⋯ | No beam | ⋯ | | 1 | UP | | |
| 24 | 18:05 | 1 | $1.55 \times 10^{11}$ | 0.94 | 1.65 | OUT | 1 | 4 |
| 25 | 18:18 | 1 | $2.00 \times 10^{11}$ | 0.94 | 1.66 | IN | 1 | 4 |
| 26 | 18:28 | 1 | $1.89 \times 10^{11}$ | 1.22 | 1.41 | IN | 1 | 4 |

Table I shows the beam parameters and data acquisition details recorded during the experiment.

The powder experiment apparatus (Fig. 1) was designed to permit the observation of proton beam interactions with an open container of tungsten powder using high speed photography and laser-Doppler vibrometry (LDV).

The tungsten powder sample was extracted from a larger batch of material that had previously been used in a series of pneumatic conveying trials [3]. The grain size distribution (Fig. 5) was evaluated using a commercially available particle size analyzer. The maximum grain size was around 250 microns (60 mesh) and the mean aerodynamic diameter was estimated to be of the order of 30 microns. Under the microscope the grain shapes appeared highly irregular and nonspherical (Fig. 6). The tapped bulk powder density was approximately 9 g/cc.

The powder container (Figs. 3 and 4) took the form of a 15 mm wide by 22 mm deep by 300 mm long *U*-shaped open topped trough, closed off at both ends, and constructed from 1 mm thick grade-II titanium sheet. The trough had a double wall such that the inner wall was in contact with the tungsten powder while the outer wall was not. This was to permit the LDV to distinguish between vibrations induced solely by secondary particle interactions and any additional vibrations generated by contact with the powder. A viewing hole in the outer trough wall generated a line of sight to the inner trough for the LDV. Approximately

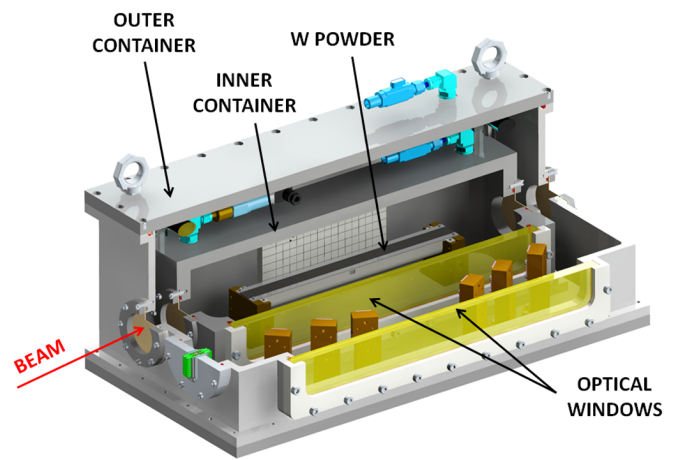

FIG. 1. Section drawing of tungsten powder experiment container.

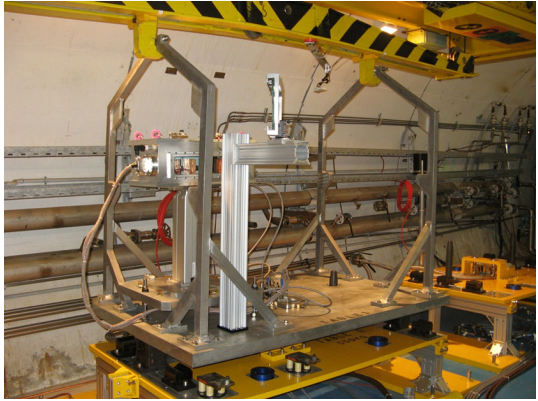

FIG. 2.   Experiment container mounted on a remotely operated table and installed in the tunnel ready for beam.

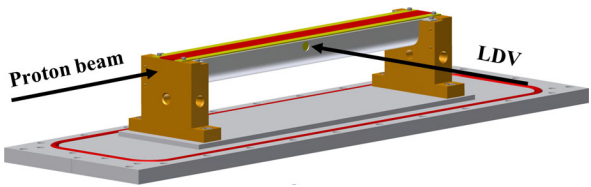

FIG. 3.   Tungsten powder trough.

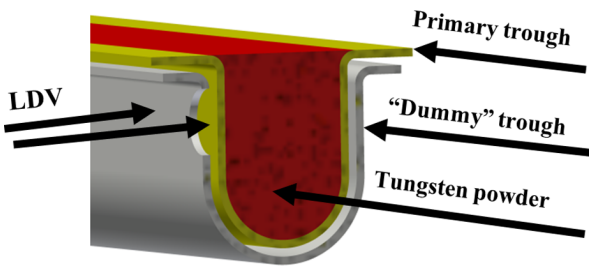

FIG. 4.   Section view of the powder trough.

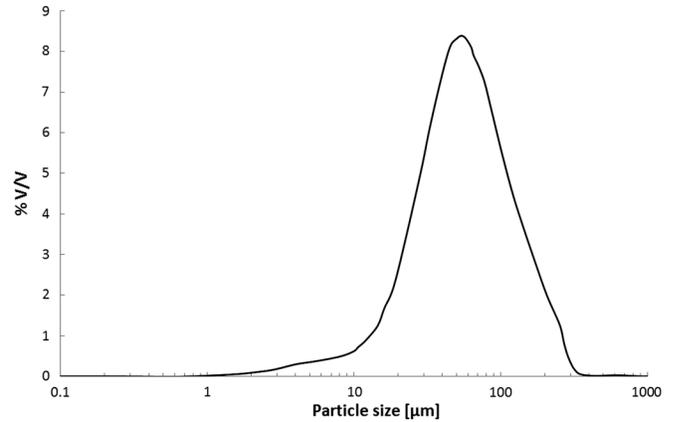

FIG. 5.   Powder size distribution percentage [% $V/V$].

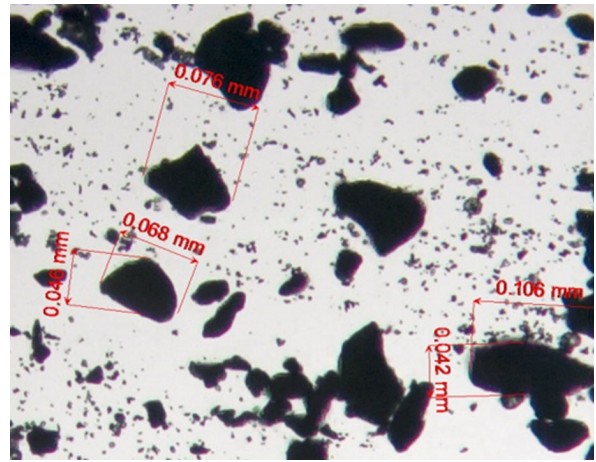

FIG. 6.   Microscope image of tungsten powder.

210 g of tungsten powder was poured into the trough and tapped (but not compacted), slightly exceeding a level fill in order to permit a direct side view of the powder top surface. The container was oriented so that the beam would intercept the sample along its length with its vertical position set starting at 6 mm below the powder top surface.

In order to mitigate the risk of airborne particulate contamination of the irradiation area a system of double containment was adopted. This comprised two hermetically sealed aluminum vessels, one mounted inside the other. The open trough was housed inside the inner vessel. 15 mm thick rectangular soda-lime glass windows mounted on the vessel sides allowed a lateral view of the trough for the camera and LDV. A grid with 1 cm vertical and horizontal line spacing was engraved on a plate mounted to the back wall of the inner container to allow direct calibration of distances and pixel sizes in the high speed photography. Upstream and downstream grade-5 titanium beam windows were mounted to the vessels at the beam entry/exit positions. On assembly, both vessels were evacuated and then back-filled to atmospheric pressure with helium to provide an inert environment for the test.

The apparatus was fastened to a vertical lift platform such that the depth of the beam below the powder surface could be adjusted. Instead, the lift table itself was mounted on a specially designed kinematic base table that allowed the apparatus to be remotely installed onto the beam line in the irradiation area (Fig. 2). The location of the target axis with respect to the theoretical beam trajectory was accurate to a precision of 0.2 mm.

The prompt radiation level precluded the placement of any electronic equipment in the near vicinity of the apparatus. Instead, the high speed camera and LDV were located in a custom made concrete bunker approximately 35 meters away from the apparatus. A line of sight onto the sample was achieved via two flat mirrors, one placed close to the bunker and the other mounted close to the apparatus.

A monochrome high speed camera manufactured by RedLake, model MotionXtra HG-100k, was used at a frame rate of 1000 and 2000 frames per second. A telephoto lens arrangement was used to focus the camera onto the apparatus. This comprised a 1000 mm fixed aperture manual focus F11 NIKKOR mirror lens coupled to a 1.4x teleconverter and a 2x teleconverter. This provided a total effective focal length of 2800 mm and a relative aperture of $f$ 32. An image resolution of $768 \times 480$ pixels, covering a field of view about $12.5 \times 7$ cm was achieved, corresponding to a pixel size of around 160 microns. The camera field of view incorporated the trough side wall as well as about 4.5 cm of open space above the trough. An example of the camera field of view is shown in Fig. 12.

To compensate for the high frame rate, remote distance and small aperture of the camera, a high intensity LED lighting rig was constructed, generating a total luminous flux of around 18 000 lumens. The lighting rig consisted of 12 OSLON LED clusters, each cluster containing 10 LEDs and being driven by a 700 mA constant current supply. The LED clusters were mounted on copper heat sinks that were placed directly outside the glass observation window of the inner containment vessel to provide "front lighting" of the sample and trough.

A laser Doppler vibrometer manufactured by Polytec, model OFV-505, was used at a sampling frequency of 10.24 MHz. The laser wavelength was 633 nm and its spot diameter at the trough was around 6 mm.

Both instruments were triggered through an electronic signal that was synchronized to the beam delivery. The camera had a pretrigger buffer of 20 frames and recorded a total of 2000 frames. The LDV recorded for a total of 12 milliseconds, of which 1.2 ms were pretrigger.

## III. FLUKA SIMULATIONS

Simulations of the energy deposition in the tungsten powder sample and its titanium container were carried out using FLUKA [8]. The tungsten powder was modeled as a compound of helium and tungsten each taking up 50% of the sample volume. The energy density in the tungsten grains was assumed to be double the value calculated in the compound based on the fact that the energy deposited in the helium is negligible compared to the energy deposited in the tungsten. Figure 7 shows the energy deposition in the tungsten powder compound as a function of radius and axial position. This indicated prior to the experiment that a peak energy deposition would be located at a longitudinal distance of 11 cm into the powder. The camera field of view was restricted to the central part of the trough covering the range 9 cm $< z < 21$ cm (Fig. 12). Figure 8 shows how the energy deposition in the tungsten and the helium converts to a temperature jump (assuming constant specific heat capacities) and indicates the large range of temperature jump that would be possible in the HiRadMat facility. Figure 9 shows a section view of the energy deposition in the sample and two troughs.

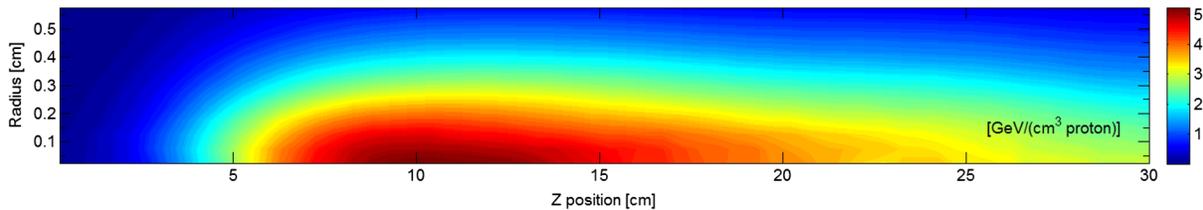

FIG. 7.   Energy deposition in the tungsten-helium compound for 440 GeV, 2 mm beam sigma.

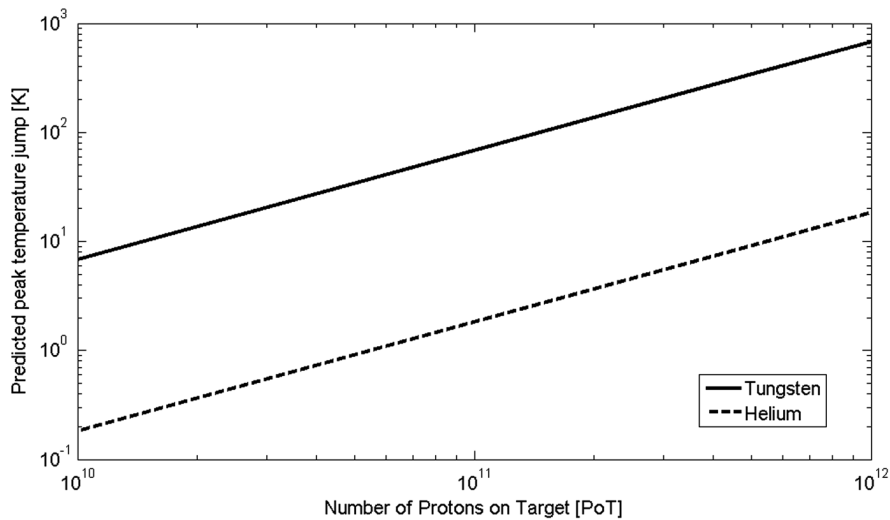

FIG. 8.   Predicted maximum temperature jump assuming constant heat capacity and beam sigma of 2 mm.

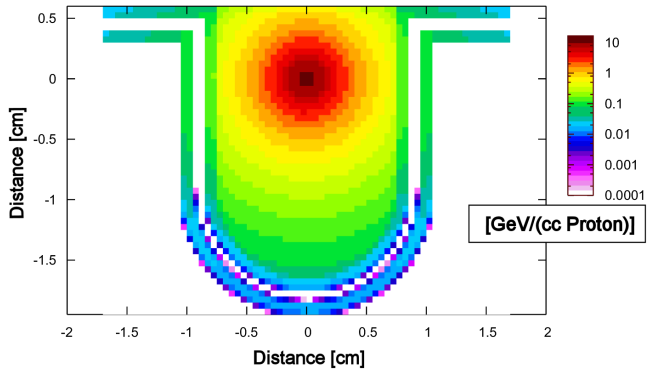

FIG. 9.   Energy deposition in sample and titanium troughs for a 440 GeV, 2 mm beam sigma.

## IV. RESULTS/OBSERVATIONS

### A. High speed camera

During the experiment the beam intensity was progressively ramped up from a minimum of $3.5 \times 10^9$ protons on target (PoT) to a maximum of $2.94 \times 10^{11}$ PoT. Videos 1–5 show the powder in the trough before and 30 ms after beam impact. Examination of the high-speed video (HSV) footage of the experiment highlighted a perceptible powder lift (Video 1) from the trough at a beam intensity of $4.6 \times 10^{10}$ (having noticed no lift at $6.8 \times 10^9$ PoT) and displayed a significant powder lift at $1.75 \times 10^{11}$ PoT (Video 2). The following experiment displayed a more dramatic and inhomogeneous powder lift at a comparable intensity of $1.85 \times 10^{11}$ PoT (Video 3). Tracking of the powder front on sequential time stamped HSV frames enables estimation of the lift velocities. The maximum powder lift velocity varied from 0.44 m/s for $1.75 \times 10^{11}$ PoT (Video 2) to 1.2 m/s for $2.9 \times 10^{11}$ PoT (Video 4). Please note that the lighting was such that a shadow was cast by the powder on the graduated back plate.

On some of the beam shots a secondary powder disruption was observed (Video 5). This occurred as the primary disruption was settling and appeared at two distinct locations approximately symmetric with respect to the trough ends.

Figure 10 shows the maximum and minimum heights reached by the powder during the lifts as a function of PoT as typified in Video 4(b). Figure 11 shows the time evolution of the powder disruption.

A clear correlation between beam intensity and lift height emerges from the results. Note however that after the first powder lift the packing and distribution of the powder on the trough is likely to have been perturbed significantly as can be seen by comparing Fig. 12 with Fig. 14. A photo taken after the experiments (Fig. 12) shows a trench in the powder sample confirming that some of the powder was ejected from the trough (Fig. 13). Such a trench may account for the difference in the behavior of the powder in subsequent shots of similar intensity [e.g., Video 2(b) compared to Video 3(b)].

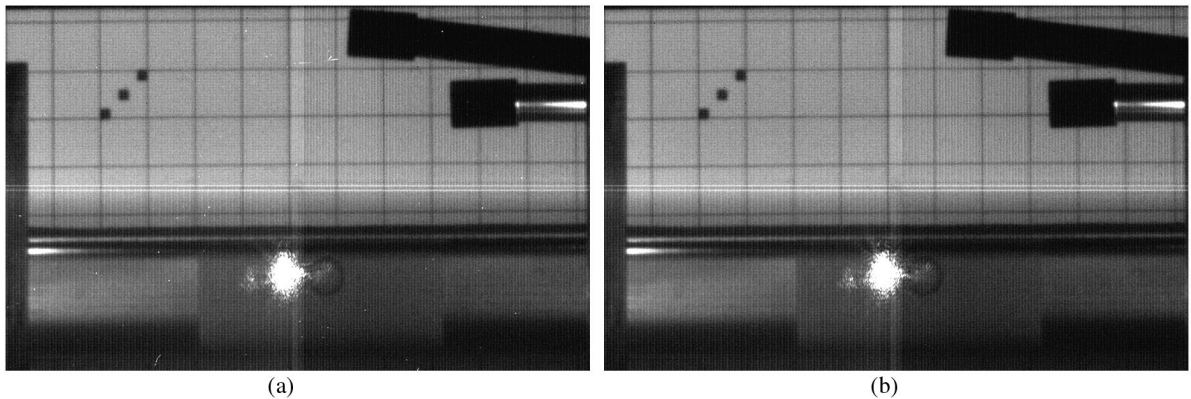

VIDEO 1.   (a) Shot 7 before proton beam pulse. (b) Shot 7, 30 ms after beam pulse.

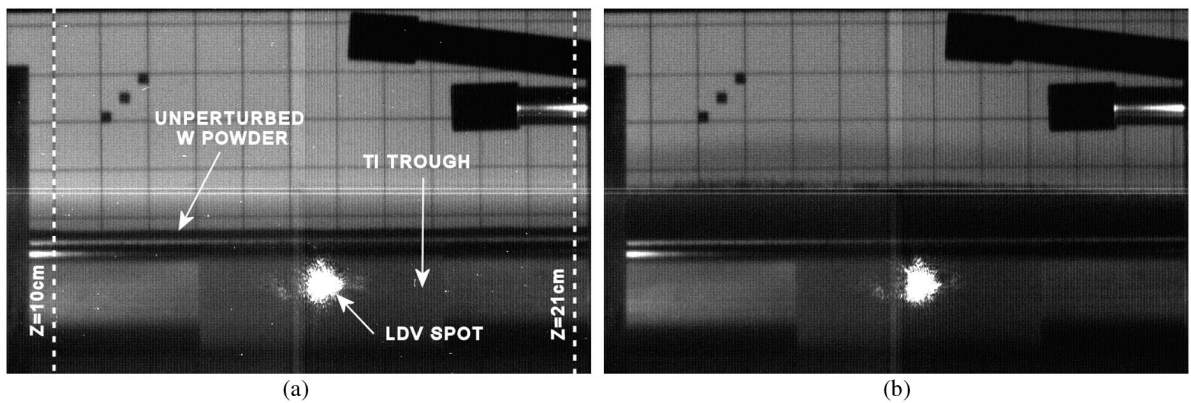

VIDEO 2.   (a) Shot 8 before proton beam pulse. (b) Shot 8, 30 ms after beam pulse (note uniform powder lift).

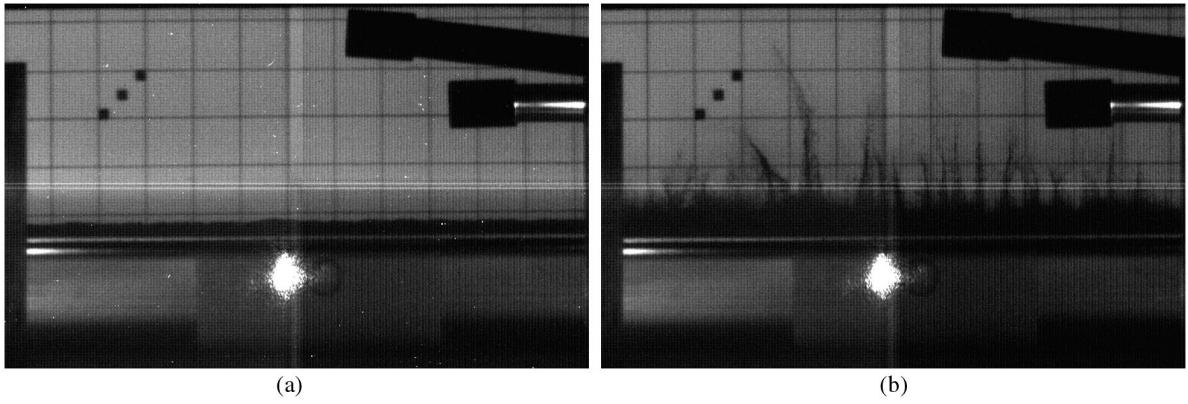

VIDEO 3.   (a) Shot 9, before beam pulse (note increased height of powder resting on trough lip resulting from previous eruption). (b) Shot 9, 30 ms after beam pulse (note nonuniform powder lift).

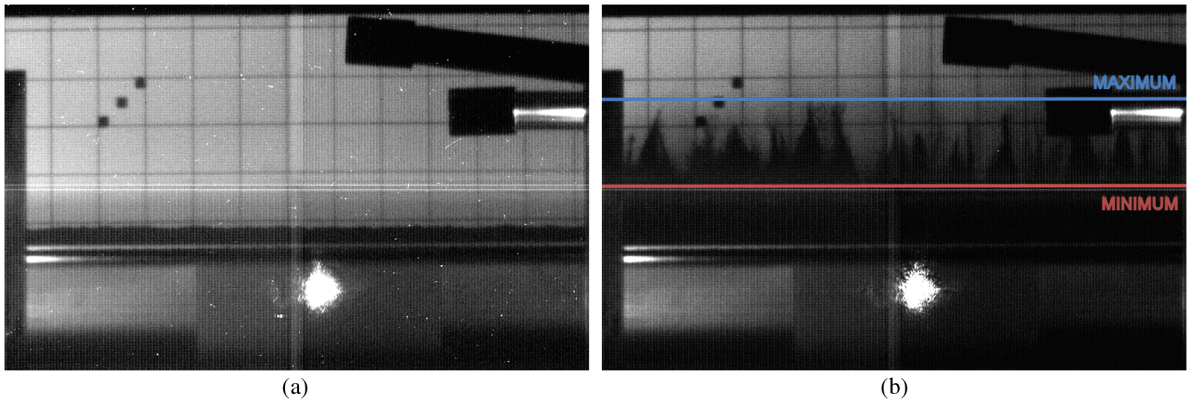

VIDEO 4.   (a) Shot 21 before beam pulse. (b) Shot 21, 30 ms after beam pulse.

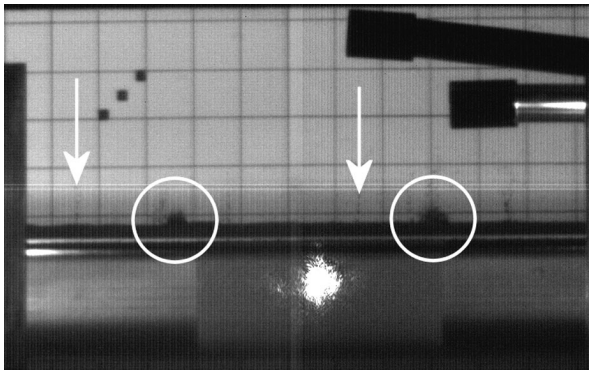

VIDEO 5.   Shot 10, 76 ms after beam pulse. The arrows highlight tails of the primary disruption settling whilst the circles highlight the two secondary disruptions rising.

## B. Laser Doppler vibrometer (LDV) data

Due to access restriction in the experimental area, the LDV setup could not be calibrated and tested *in situ* prior to the experiments. Analysis of the LDV measurements identified a variable delay between the beam trigger and the data recording. This made it difficult to accurately resolve the propagation speed of the sonic wave from beam impact through the powder. This should be resolved in a future experiment.

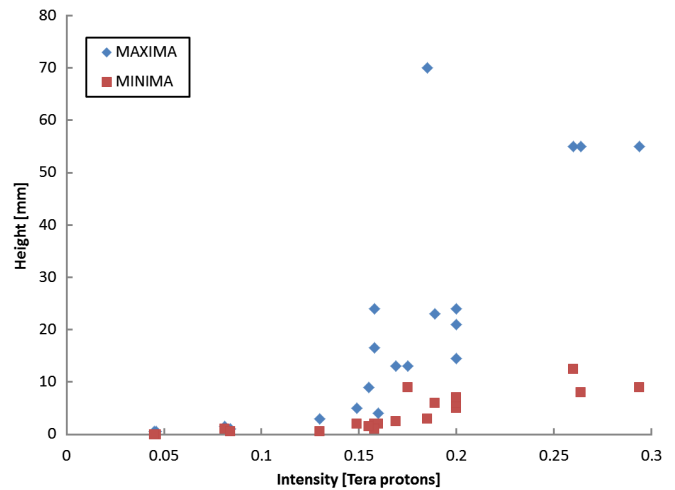

FIG. 10.   Relationship between intensity and height of powder lift in all beam shots.

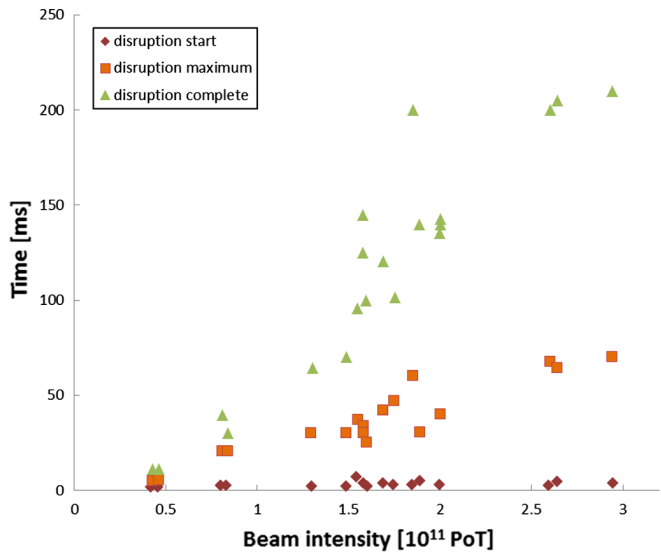

FIG. 11.   Time evolution of the powder disruption.

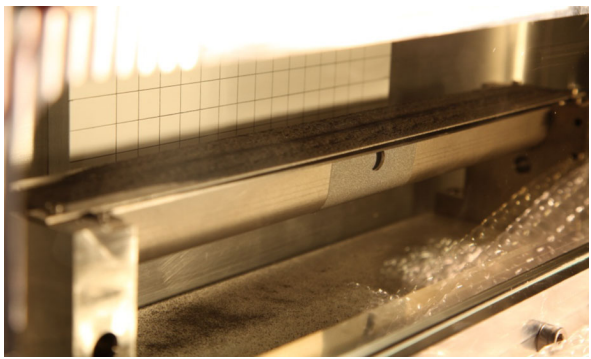

FIG. 12.   Photo of powder trough after the experiment showing deep trench made in the powder sample and powder on the base of the containment box.

Analysis of the LDV data also revealed a high level of intermittent noise. When filtered to account for these factors, the LDV signals showed a dominant vibration at around 1 kHz. Simple modal analysis of an empty trough like the one used in the experiment confirms 1 kHz to be a resonant frequency for the structure.

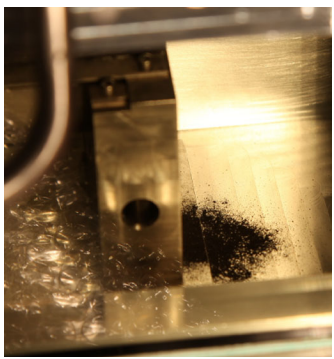

FIG. 13.   Photo of the powder spilled out from the end of the trough.

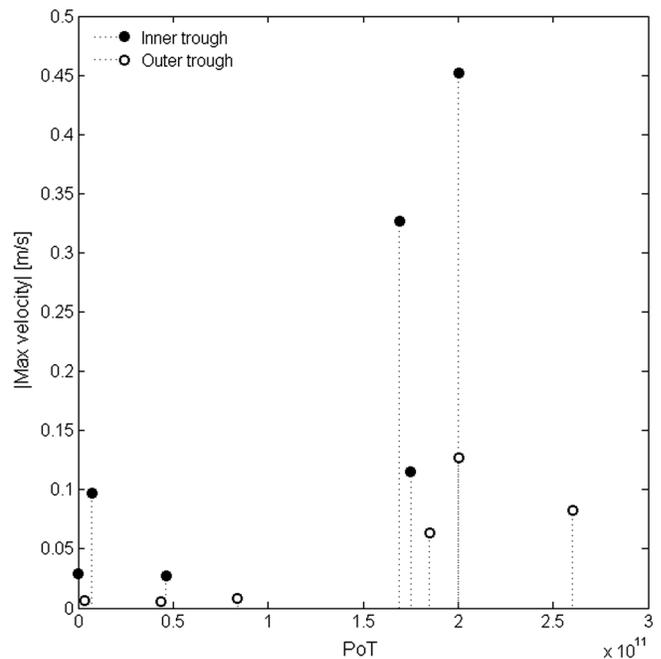

FIG. 14.   LDV data of trough surface velocity as a function of beam intensity.

As shown in Fig. 14 the signal amplitude measured on the inner trough was found consistently greater than that acquired on the outer trough (Figs. 4 and 9). The data also highlights a proportionality between the signal amplitude and the beam intensity.

## V. CONCLUSIONS

In-beam experiments demonstrated that a sample of tungsten powder immersed in a helium atmosphere is perturbed when impinged by a 440 GeV proton beam with a threshold intensity of around $4.6 \times 10^{10}$ protons and a horizontal and vertical beam sigma of 0.45 mm and 1 mm respectively. This intensity threshold corresponds to a peak energy density of approximately 7.5 J/g in the tungsten grains.

The experiment reached a similar pulsed energy density as the mercury trough experiment, i.e., approximately 4 or 5 times lower than the energy density expected in a neutrino factory target. The powder eruption velocity was found to be 68 times lower than the previously reported splash velocities for mercury subjected to an equivalent pulsed energy density, i.e., 0.44 m/s at 29 J/g for tungsten powder compared with 30 m/s at 28 J/g for mercury. Lower perturbation velocities are likely to have a less damaging impact on the target containment infrastructure.

The maximum velocity recorded was 1.2 m/s at $2.9 \times 10^{11}$ PoT.

## ACKNOWLEDGMENTS

The authors would like to thank the HiRadMat operation team for the technical support during the experiment. The research leading to these results has received partial funding from the European Commission and Science and Technology Facilities Council (STFC) under the FP7 Capacities project EuCARD, Grant Agreement No. 227579.

———————————

[1] T. W. Davies, O. Caretta, C. J. Densham, and R. Woods, Powder Technol. **201**, 296 (2010).

[2] S. Choubey *et al.*, Technical Report No. FERMILAB-DESIGN-2011-01, http://www.osti.gov/scitech/biblio/1029650.

[3] O. Caretta *et al.*, in *Proceedings of the 11th European Particle Accelerator Conference, Genoa, 2008* (EPS-AG, Genoa, Italy, 2008), WEPP161.

[4] I. Efthymiopoulos *et al.*, *Proceedings of the 2nd International Particle Accelerator Conference, IPAC-2011, San Sebastián, Spain* (EPS_AG, Spain, 2011), TUPS058 [http://accelconf.web.cern.ch/AccelConf/IPAC2011/papers/tups058.pdf].

[5] J. Back, C. Densham, R. Edgecock, and G. Prior, Phys. Rev. ST Accel. Beams **16**, 021001 (2013).

[6] J. Lettry, A. Fabich, S. Gilardoni, M. Benedikt, M. Farhat, and E. Robert, J. Phys. G **29**, 1621 (2003).

[7] I. Efthymiopoulos *et al.*, in *Proceedings of the 23rd Particle Accelerator Conference, Vancouver, Canada, 2009* (IEEE, Piscataway, NJ, 2009), TU6PFP085.

[8] A. Fasso, A. Ferrari, J. Ranft, and P. R. Sala, *FLUKA: A multiparticle transport code, 2005.*